\documentclass[showpacs,english,prd,preprint,nofootinbib,preprintnumbers]{revtex4}

\usepackage{graphicx}
\usepackage{amsmath}
\usepackage{amssymb}
\usepackage{bm}
\usepackage{latexsym}
\usepackage{epsfig}
\usepackage{verbatim}

\begin{document}
\title{Cosmological and Solar-System Tests of $f(R)$ Modified Gravity}
\date{\today}

\author{Wei-Ting Lin$^{1,2}$}
\email{r97222021@ntu.edu.tw}
\author{Je-An Gu$^{2}$}
\email{jagu@ntu.edu.tw}
\author{Pisin Chen$^{1,2,3,4}$}
\email{pisinchen@phys.ntu.edu.tw}
\affiliation{
${}^1$Department of Physics, National Taiwan University, Taipei 10617, Taiwan, R.O.C.\\
${}^2$Leung Center for Cosmology and Particle Astrophysics, National Taiwan University, Taipei 10617, Taiwan, R.O.C.\\
${}^3$Graduate Institute of Astrophysics, National Taiwan University, Taipei 10617, Taiwan, R.O.C.\\
${}^4$Kavli Institute for Particle Astrophysics and Cosmology, SLAC National Accelerator Laboratory, Menlo Park, CA 94025, U.S.A.
}

\begin{abstract}
We investigate the cosmological and the local tests of the
$f(R)$ theory of modified gravity via the observations of
(1) the cosmic expansion and (2) the cosmic structures and
via (3) the solar-system experiments.
To fit the possible cosmic expansion histories under
consideration, for each of them we reconstruct $f(R)$, known as
``designer $f(R)$''. We then test the designer $f(R)$ via the
cosmic-structure constraints on the metric perturbation ratio
$\Psi/\Phi$ and the effective gravitational coupling
$G_\textrm{eff}$ and via the solar-system constraints on the
Brans-Dicke theory with the chameleon mechanism. We find that
among the designer $f(R)$ models specified by the CPL effective
equation of state $w_\textrm{eff}$, only the model closely
mimicking general relativity with a cosmological constant
($\Lambda$CDM) can survive all the tests. Accordingly, these
tests rule out the frequently studied $w_\textrm{eff}=-1$
designer $f(R)$ models which are distinct in cosmic structures
from $\Lambda$CDM. When considering only the cosmological
tests, we find that the surviving designer $f(R)$ models,
although exist for a variety of $w_\textrm{eff}$, entail
fine-tuning.
\end{abstract}


\maketitle

\section{Introduction}
\label{introduction}

Our universe is presently at the stage of accelerating
expansion, as suggested by the cosmological observations
\cite{Perlmutter:1998np,Riess:1998cb}. Assuming a homogeneous
and isotropic universe and based on general relativity (GR),
one cannot explain this late-time cosmic acceleration if the
universe contains only matter and radiation. A substance with
negative pressure adding to the right-hand side of the Einstein
field equations, known as dark energy \cite{Huterer:1998qv},
can lead to the cosmic acceleration \cite{DEreview}.
Observations
\cite{Komatsu:2010fb,Kowalski:2008ez,Eisenstein:2005su} suggest
that our universe should be nearly flat and consist of 73\%
dark energy, 22\% dark matter, 5\% ordinary matter, and
approximated 0.008\% radiation. It is ironic that so far our
knowledge of physics can only explain 5\% of the universe, with
nearly 95\% of the universe remained unexplained. This is the
most challenging issue for the physicists of the 21st century.

Instead of invoking dark energy, there have been attempts to
modify the left-hand side of the Einstein equations, e.g.\ via
proposing alternative gravity theories, to explain the
late-time acceleration
\cite{Dvali:2000hr,Maartens:2003tw,Bekenstein:2004ne,Jacobson:2000xp}
(see \cite{Nojiri:2006ri} for a review). These attempts are
known as modified gravity. Among these theories, in the present
paper we will focus on the $f(R)$ theory of modified gravity
where the Einstein-Hilbert Lagrangian density is added by a
general function of the Ricci scalar, $f(R)$.

A.~A.~Starobinsky was the first to point out that de Sitter
space, and therefore the exponential expansion of spacetime,
can be achieved in the early universe by quantum correction,
($R+\alpha^2R^2$), to the Einstein-Hilbert action
\cite{Starobinsky:1980te}. This UV modification leads to an
early accelerating universe (Starobinsky inflation) and
motivates people to construct the $f(R)$ models
\cite{Capozziello:2003tk,Carroll:2003wy} to explain the
late-time acceleration of the universe. Since the $f(R)$ theory
is equivalent to the Brans-Dicke theory with the Brans-Dicke
parameter $\omega_\textrm{BD}=0$ \cite{Chiba:2003ir}, it seems
that this is already ruled out by the constraint
$\omega_\textrm{BD}>4\times 10^4$ set by observations
\cite{Will:2001mx,Bertotti:2003rm,Will:2005va}.
Nevertheless, it was pointed out recently that in a
non-minimally coupled scalar-tensor theory the scalar field may
have a scale- or environment-dependent mass via its interaction
with matter \cite{Khoury:2003aq,Khoury:2003rn}. That is, the
higher (lower) the matter density is, the more (less) massive
the scalar field is. Consequently, in the solar system, which
has a much higher density than the universe, the scalar field
would have a large effective mass (and therefore a small
Compton wavelength)
so that it can evade the solar-system tests. At the mean time,
the same scalar field would have a much smaller effective mass
at cosmological scales, as small as the present Hubble
expansion rate (i.e.\ a long Compton wavelength comparable to
the size of the universe). Accordingly, it can be a candidate
driving the late-time cosmic acceleration. The sensitivity to
the environment of such a scalar field interaction gives the
name, ``chameleon mechanism''
\cite{Khoury:2003aq,Khoury:2003rn}.



In this paper we investigate various tests of the $f(R)$ theory
of modified gravity. As an alternative to dark energy for
driving the late-time cosmic acceleration, the $f(R)$ theory
needs to pass the cosmological tests, including the constraints
about (1) the cosmic expansion and (2) the cosmic structure
formation. In addition, as a modified gravity theory, it needs
to pass (3) the solar-system test, such as the constraints on
the Brans-Dicke theory. The process of these three tests in the
present paper are described in the following.


\textbf{(1) Cosmic Expansion History:} %
For the background cosmic expansion there is a degeneracy
between $f(R)$ modified gravity and GR with dark energy, that
is, one cannot distinguish between these two theories via the
observations about the cosmic expansion history such as type Ia
supernovae (SNIa) and baryon acoustic oscillations (BAO). This
is because for any given expansion history $H(t)$
one can always construct a function $f(R)$ to generate the
required expansion. The $f(R)$ model so constructed is called
``designer $f(R)$'' \cite{Pogosian:2007sw}. To fit the
observational constraint on $H(t)$, we take the designer $f(R)$
approach to (re)construct $f(R)$ for possible expansion
histories, where $f(R)$ is specified by $H(t)$ and the initial
condition, particularly the value of $df/dR$ at some initial
time, as denoted by $f_{Ri}$. In the present paper we consider
the expansion history $H(t)$ given by the effective dark energy
equation of state $w_\textrm{eff}$ as specified by the
Chevallier-Polarski-Linder (CPL) parametrization
\cite{Chevallier:2000qy,Linder:2002et}, $w_\textrm{CPL} = w_0 +
w_a z/(1+z)$ (where $z$ is redshift), with the current
observational constraints on $w_0$ and $w_a$
\cite{Komatsu:2010fb}.
In this case the designer $f(R)$ is parametrized by three
parameters, $\{ w_0,w_a,f_{Ri} \}$. We then test such designer
$f(R;\{w_0,w_a,f_{Ri}\})$ model via the cosmic-structure
observations and the solar-system experiments.

\textbf{(2) Cosmic Structure Formation:} %
For each designer $f(R;\{w_0,w_a,f_{Ri}\})$ obtained above we
calculate the theoretical prediction of the cosmological
perturbations, particularly $\Psi/\Phi$ and
$G_\textrm{eff}/G_{N}$ at late times and at subhorizon scales,
where $\Psi$ and $\Phi$ are two metric potentials in the
Newtonian gauge, and $G_\textrm{eff}$ and $G_{N}$ are the
effective gravitational coupling and the Newtonian constant,
respectively. In GR these two ratios are both unity, while in
$f(R)$ modified gravity they generally change with time and
scales. For the test we compare the theoretical prediction with
the observational requirements from the cosmic structure
observations, thereby constraining $f(R;\{w_0,w_a,f_{Ri}\})$,
i.e., obtaining the constraints on the parameter space $\{ w_0
, w_a , f_{Ri} \}$. In particular, we obtain the allowed range
of $f_{Ri}$ for various $\{ w_0 , w_a \}$ under consideration.

\textbf{(3) Solar-system Tests:} %
The solar-system experiments provide stringent constraints on
the Brans-Dicke theory with the chameleon mechanism, and
accordingly tightly constrain the behavior of $f(R)$ for
$R/H_0^2 \gtrsim 10^5$
\cite{Faulkner:2006ub,Capozziello:2007eu,Bisabr:2009ee,Gu:2010}.
To do the solar-system test, for each designer
$f(R;\{w_0,w_a,f_{Ri}\})$ we check whether it is consistent
with this constraint on $f(R/H_0^2\gtrsim10^5)$.



Through the above three tests we find that only the models
which closely mimic GR with a cosmological constant can survive
these three tests. That is, among the $f(R;\{w_0,w_a,f_{Ri}\})$
models under consideration, only the case where $f(R) \simeq
\textrm{constant}$ or $\{w_0,w_a,f_{Ri}\}$ very close to
$\{-1,0,0\}$ survives these three tests. For comparison, when
we consider only the cosmological tests, i.e.\ the tests (1)
and (2), for each $\{w_0,w_a\}$ under consideration we find the
allowed range of $f_{Ri}$ very narrow although existing. This
presents a severe fine-tuning of the initial condition of
$f(R)$ for passing the cosmological tests.


This paper is organized as follows. In Sec.\ \ref{background}
we introduce the basics of the background evolution in the
$f(R)$ theory, the construction of the designer $f(R)$ models,
and the evolution of the cosmological perturbations with the
late-time and the subhorizon approximations. In Sec.\ \ref{obs}
we introduce the observational constraints set by the
observations about the cosmic expansion and the cosmic
structures and by the solar-system experiments. In Sec.\
\ref{results} we show the results of the three tests. We then
conclude in Sec.\ \ref{conclusion}.

\section{Basics of $f(R)$ Modified Gravity}
\label{background}
\subsection{Background evolution}
\label{backgroundbasics}

We start with the following modified Hilbert-Einstein action
for the $f(R)$ theory of modified
gravity.\footnote{$\kappa^2\equiv8\pi G$}
\begin{equation}
S=\frac{1}{2\kappa^{2}}\int d^4 x\sqrt{-g}~
\left[R+f(R)\right]+\sum_{\scriptstyle{a}} S_{a} \ ,
\label{f(R)action}
\end{equation}
where $f(R)$ is a general function of the Ricci scalar and
represents the deviation from GR, and $S_{a}$ with $a=r,m$ are
the actions for radiation and matter (including dark matter and
baryons), respectively. The variation of this action with
respect to the metric tensor gives the gravitational field
equations,
\begin{eqnarray}
&&G_{\mu\nu}+\left(f_RR_{\mu\nu}-\frac{1}{2}fg_{\mu\nu}\right)+\left(g_{\mu\nu}\square-\nabla_\mu\nabla_\nu\right)f_R\nonumber\\
&&=\kappa^2T_{\mu\nu}=\kappa^2\sum_{\scriptstyle{a}}T^{(a)}_{\mu\nu}\ ,
\label{f(R)fieldequation}
\end{eqnarray}
where the d'Alembertian operator
$\square\equiv\nabla_\nu\nabla^\nu$, $f_R\equiv df/dR$,
likewise $f_{RR}\equiv d^2f/dR^2$, and the energy-momentum
tensor of the cosmic fluid $T^{(a)}_{\mu\nu} \equiv
-(2/\sqrt{-g}) (\delta S^{(a)} / \delta g^{\mu\nu})$.

Regarding the background evolution, we consider the flat
Robertson-Walker metric for describing the background
space-time,
\begin{equation}
ds^2=-dt^2+a^2(t)d\vec{x}~^2\ ,
\label{backmetric}
\end{equation}
and the perfect fluids for the energy contents,
\begin{equation}
T^{\mu}_{~\,\nu}=\sum_{a} diag(-\rho_a,p_a,p_a,p_a) ,
\label{backtmunu}
\end{equation}
where $\rho_{a}$ and $p_{a}$ are the energy density and pressure.
%
We then obtain two evolution equations describing the cosmic
expansion:
\begin{eqnarray}
H^2+\frac{1}{6}f+\left(H^2-\frac{1}{6}R\right)f_R+H\dot{f_R}
&=& \frac{\kappa^2}{3}\left(\rho_m+\rho_r\right) ,
\label{eff1Friedmann} \\
\frac{\ddot{a}}{a}+\frac{1}{6}f-H^2f_R+\frac{1}{2}H\dot{f_R}+\frac{1}{2}\ddot{f_R}
&=& \frac{-\kappa^2}{6}\left(\rho_m+2\rho_r\right) ,
\label{eff2Friedmann}
\end{eqnarray}
where an over-dot denotes the derivative with respect to the
cosmic time $t$, the Hubble expansion rate $H=\dot{a}/a$, the
Ricci scalar $R=6\dot{H}+12H^2$, and we have used $p_m=0$ and
$p_r=\rho_r/3$. Equation (\ref{eff2Friedmann}) shows that
matter and radiation can simply contribute to deceleration,
while the deviation from GR may induce accelerated expansion
when its contribution therein,
$f/6-H^2f_R+H\dot{f_R}/2+\ddot{f_R}/2$, is negative and
significant. In addition, from these equations one can see that
the $f(R)$ theory comes back to GR (with a cosmological
constant $\Lambda$) at the background level when
$f(R)=\textrm{constant}=-2\Lambda$.

The above two equations can be recast in the form in GR:
\begin{eqnarray}
H^2 &=& \frac{\kappa^2}{3}\left(\rho_m+\rho_r+\rho_\textrm{eff}\right) ,
\label{effFriedmann1designer} \\
\frac{\ddot{a}}{a} &=& -\frac{\kappa^2}{6}\left(\rho_m+2\rho_r+\rho_\textrm{eff}+3p_\textrm{eff}\right) .
\label{effFriedmann2designer}
\end{eqnarray}
Here $\rho_\textrm{eff}$ and $p_\textrm{eff}$ are respectively
the effective energy density and the effective pressure
contributed from the deviation from GR, $f(R)$, and are defined
as
\begin{eqnarray}
\rho_\textrm{eff} &\equiv& \frac{1}{\kappa^2} \left[ -\frac{1}{2}f -3\left(H^2-\frac{1}{6}R\right)f_R -3H\dot{f_R} \right] ,
\label{effrho} \\
p_\textrm{eff} &\equiv& \frac{1}{\kappa^2} \left[ \frac{1}{2}f -\left(H^2+\frac{R}{6}\right)f_R +2H\dot{f_R}+\ddot{f_R} \right] .
\label{effPressure}
\end{eqnarray}
This effective energy density satisfies the conservation
equation,
\begin{equation}
\dot{\rho}_\textrm{eff}=-3H\left(1+w_\textrm{eff}\right)\rho_\textrm{eff} \, ,
\label{effcontinuity}
\end{equation}
where
\begin{equation}
w_\textrm{eff} \equiv p_\textrm{eff} / \rho_\textrm{eff} \, .
\label{effw}
\end{equation}

For a given expansion history $H(t)$ and given energy
densities, $\rho_m(t)$ and $\rho_r(t)$, Eq.\
(\ref{eff1Friedmann}) gives a second-order differential
equation for $f(R)$ or $f(t)$, with the understanding that
$f_R=\dot{f}/\dot{R}$ and $\dot{f}_R = \ddot{f}/\dot{R} -
\ddot{R}\dot{f}/\dot{R}^2$. Equivalently, for a given effective
energy density $\rho_\textrm{eff}$ that together with $\rho_m$
and $\rho_r$ specifies the expansion history, Eq.\
(\ref{effrho}) gives the same differential equation for $f(R)$
\cite{Pogosian:2007sw}:
\begin{eqnarray}
\lefteqn{ \frac{f''}{H^2_0}-\left(1+\frac{E'}{2E}
+\frac{R''}{R'}\right)\frac{f'}{H^2_0}} \nonumber \\
&& +\left(\frac{12E'}{E}+\frac{3E''}{E}\right) \left(\frac{f}{6H^2_0}+E_\textrm{eff}\right) =0 ,
\label{odeforf(R)}
\end{eqnarray}
where the prime denotes the derivative with respect to $\ln a$,
and
\begin{eqnarray}
E(a) &\equiv& H^2/H^2_0 = \Omega_m a^{-3} + \Omega_r a^{-4} + E_\textrm{eff} , \\
E_\textrm{eff}(a) &\equiv& \rho_\textrm{eff} / \rho_c
= \Omega_\textrm{eff} \exp \left\{ -3 \int_0^{\ln a}
\left[ 1+w_\textrm{eff}(\tilde{a}) \right] d\ln \tilde{a} \right\} .
\end{eqnarray}
Here the density fraction $\Omega_i \equiv \rho_i
(\textrm{now})/\rho_c$ (for $i = m,r,\textrm{eff}$), $\rho_c$
is the present critical density, and
$\Omega_{m} + \Omega_r + \Omega_\textrm{eff} = 1$
under our consideration of a flat universe. %
Accordingly, the function $f(R)$ which satisfies this
differential equation can generate the required expansion
history. Since in general the solution of this differential
equation exists (and is not unique), for any given expansion
history there always exists a function $f(R)$ which can
generate that required expansion. As a result, we cannot
distinguish between $f(R)$ modified gravity and dark energy via
the observations about the cosmic expansion.

\subsection{Designer $f(R)$}
\label{backgrounddesignerfr}

In the study of $f(R)$ modified gravity, one may start with a
carefully chosen function $f(R)$. On the other hand, one may
instead reconstruct $f(R)$ from a given expansion history
(together with the information about the energy contents),
i.e., considering the solutions $f$ of Eq.\ (\ref{odeforf(R)})
for given cosmological parameters $\{ H_0 , \Omega_m , \Omega_r
, \Omega_\textrm{eff} \}$ and given $w_\textrm{eff}(a)$. The
$f(R)$ so constructed is called ``designer $f(R)$''
\cite{Pogosian:2007sw}. This phenomenological approach makes
sense particularly when one expects that the future
cosmological observations will give precise and detailed
information about the cosmic expansion.

Solving Eq.\ (\ref{odeforf(R)}) requires initial conditions,
with different setting of which one obtains different solution
$f(R)$. Thus, the designer $f(R)$ is in general a functional of
$\{ H_0 , \Omega_m , \Omega_r , \Omega_\textrm{eff} \}$,
$w_\textrm{eff}(a)$ and the initial conditions. In the present
paper we will fix $\{ H_0 , \Omega_m , \Omega_r ,
\Omega_\textrm{eff} \}$, while consider a variety of
$w_\textrm{eff}(a)$ and initial conditions for possible $f(R)$.

It has been shown \cite{Pogosian:2007sw} that the two initial
conditions originally needed for solving the second-order
differential equation (\ref{odeforf(R)}) can be simplified to
only one condition. This is done as follows. At the early times
we require the deviation for GR be tiny in order to satisfy the
stringent constraints about cosmic microwave background (CMB)
and big-bang nucleosynthesis (BBN), and therefore
$\rho_\textrm{eff} \ll \rho_{r,m}$. With this approximation,
one can obtain an analytic solution of Eq.\ (\ref{odeforf(R)}),
which is the sum of a particular solution and a growing-mode
and a decaying mode homogeneous solutions with two arbitrary
constants to be determined by two initial conditions. Since the
decaying mode entails a large deviation from GR at early times
and cannot fit the above-mentioned requirement, one may exclude
the decaying-mode homogeneous solution. Then only one arbitrary
constant remains and only one initial condition is needed. The
initial conditions for $f$ and $f'$ at an initial time $a_i$
can be specified as follows \cite{Pogosian:2007sw}.
\begin{eqnarray}
\label{initialconditionf}
\frac{f_i}{H^2_0} &\simeq& A_{+}a_i^{p_{+}}+A_pE_\textrm{eff}(a_i) \, , \\
\label{initialconditionfprime}
\frac{f'_i}{H^2_0} &\simeq& p_{+}A_{+}a_i^{p_{+}}-3\left[1+w_\textrm{eff}(a_i)\right]A_pE_\textrm{eff}(a_i) \, ,
\end{eqnarray}
where
\begin{eqnarray*}
A_p&=&\frac{-6c}{-3w'_\textrm{eff}(a_i)+9w^2_\textrm{eff}(a_i)+\left(18-3b\right)w_\textrm{eff}(a_i)+9-3b+c} \, , \nonumber \\
p_{+}&=&\frac{-b+\sqrt{\,b^2-4c}}{2} \, , \nonumber \\
b&\equiv&\frac{7+8r}{2\left(1+r\right)}~,~c\equiv\frac{-3}{2\left(1+r\right)} \, , \nonumber \\
r&\equiv&\frac{a_{eq}}{a_i}=\frac{1}{a_i}\frac{\Omega_{r}}{\Omega_{m}} \, .\nonumber\
\end{eqnarray*}
The remaining one arbitrary constant is the amplitude of the
growing-mode homogeneous solution, $A_{+}$. Instead of
specifying the value of $A_{+}$, one may specify the more
physical initial condition $f_R(a_i)$ that gives $f_i$ and
$f'_i$ through the relation $f_R = f' / R'$, Eqs.\
(\ref{initialconditionf}) and (\ref{initialconditionfprime}).


\subsection{Cosmological perturbations}
\label{structure}

As pointed out at the end of Sec.\ \ref{backgroundbasics},
$f(R)$ modified gravity and dark energy can lead to the same
evolution of the background space-time, and therefore we cannot
distinguish these two theories of cosmic acceleration via the
observations about the cosmic expansion history. On the
contrary, even for those leading to the same cosmic expansion,
these two theories may have distinct effects on the cosmic
structure formation, which accordingly gives an opportunity of
breaking the degeneracy.

For the cosmic structures at late times, the relevant physical
quantities we consider here are the sub-horizon modes of the
matter density perturbations $\delta_m$, and the scalar metric
perturbations in the conformal Newtonian gauge, $\Psi$ and
$\Phi$, defined by
\begin{equation}
ds^2=-\left[1+2\Psi(\vec{x},t)\right]dt^2
+ a^2\left[1-2\Phi(\vec{x},t)\right]d\vec{x}^2 \, .
\label{perturmetric}
\end{equation}
These two potentials respectively characterize the particle
acceleration, $\ddot{x}=-\nabla\Psi$, and the gravitational
potential, $\nabla^2\Phi=4\pi G_N\rho$, in the Newtonian
gravity. We consider the Fourier transformation of these three
perturbed quantities, and in the remaining of the paper
$\delta_m$, $\Psi$ and $\Phi$ will denote the Fourier modes in
$k$-space with the understanding that they are functions of
time $t$ and wave-number $k$.

With the sub-horizon and the late-time approximations, we have
the following equations for the evolution of $\delta_m$ and the
relations between $\delta_m$, $\Psi$ and $\Phi$
\cite{Tsujikawa:2007gd}.
\begin{equation}
\delta''_m + \left(2+\frac{H'}{H}\right)\delta'_m - \frac{4\pi G_\textrm{eff}\rho_m}{H^2}\delta_m \simeq 0 \, ,
\label{finaldeltamevolution}
\end{equation}
\begin{equation}
\Psi \simeq -4 \pi \left( a^2 / k^2 \right) G_\textrm{eff} \rho_m \delta_m \, ,
\label{finalpsideltam}
\end{equation}
\begin{equation}
\frac{\Psi}{\Phi} \simeq \frac{1+\frac{4k^2}{a^2}\frac{f_{RR}}{1+f_R}}{1+\frac{2k^2}{a^2}\frac{f_{RR}}{1+f_R}} \, ,
\label{finalpsiphi}
\end{equation}
where the effective gravitational coupling strength
$G_\textrm{eff}$ is given as
\begin{equation}
\frac{G_\textrm{eff}}{G_N}=\frac{1}{1+f_R}\frac{1+\frac{4k^2}{a^2}\frac{f_{RR}}{1+f_R}}{1+\frac{3k^2}{a^2}\frac{f_{RR}}{1+f_R}} \, .
\label{finalgeff}
\end{equation}
Here $G_N$ is the Newtonian gravitational constant measured in
the solar system, which is very close to $\kappa^2 / 8\pi$
because the deviation from GR in the solar system must be very
tiny (if nonzero), i.e., $|f_R| \ll 1$.

The two ratios, $\Psi / \Phi$ and $G_\textrm{eff} / G_N$, give
the key to differentiating modified gravity from GR. In GR both
of them are unity, while in $f(R)$ modified gravity they are
generally time- and scale-dependent. Equations
(\ref{finalpsiphi}) and (\ref{finalgeff}) give the theoretical
prediction of these two ratios for a given $f(R)$, which can be
conveniently used for the comparison with the existing
observational constraints to be presented in the next section.


\section{Observational Constraints}
\label{obs}

In this section we present the observational constraints about
(1) the cosmic expansion history, (2) the cosmic structure
formation, and (3) the deviation from GR in the solar system.


\subsection{Cosmic expansion}

In regard of the effects on the cosmic expansion, a $f(R)$
model is equivalent to a dark energy model with the effective
equation of state $w_\textrm{eff}$ given by Eqs.\
(\ref{effrho}), (\ref{effPressure}) and (\ref{effw}).
Accordingly, the constraints on the equation of state of dark
energy from the observations about the cosmic expansion also
constrain $f(R)$ or, more precisely, the designer $f(R)$
models. In the present paper we consider the 
constraints \cite{Komatsu:2010fb} on a constant
$w_\textrm{eff}$
and on the CPL parametrization $w_\textrm{CPL} = w_0 + w_a z /
(1+z)$:
\begin{equation} \label{wconst-constraint}
w_\textrm{eff} = \mbox{ constant } = -0.980 \pm 0.053 \
(68\% \mbox{ CL}) \, .
\end{equation}
\begin{equation} \label{wCPL-constraint}
w_{0} = -0.93 \pm 0.13 \, , \; %
w_{a} = -0.41^{+0.72}_{-0.71} \ (68\% \mbox{ CL}) \, .
\end{equation}

We will consider the designer $f(R)$ models which are
constructed with respect to a variety of $w_\textrm{eff}$
following the above constraints. Such designer $f(R)$ models by
construction can pass the cosmic-expansion test.


\subsection{Cosmic structure formation}
\label{intlarge}

Recently many works on the test of modified gravity via the
observations about the cosmic structures have been done
\cite{Serra:2009kp,Daniel:2009kr,Guzik:2009cm,Giannantonio:2009gi,Daniel:2010ky,Bean:2010zq}.
In the present paper, for the test of $f(R)$ modified gravity
we will invoke the observational constraints on $\Psi/\Phi$ and
$G_\textrm{eff}/G_N$ obtained in \cite{Giannantonio:2009gi}
from the observational data about the integrated Sachs-Wolfe
(ISW) effect, which is measured by cross-correlating the CMB
with the galaxy data (the tracer of the large-scale structure)
with the help of the type Ia supernova (SN Ia) data for the
information about the cosmic expansion.

In \cite{Giannantonio:2009gi} the cosmic background expansion
is fitted with the $\Lambda$CDM model, and the cosmic
perturbations $\Psi/\Phi$ and $G_\textrm{eff}/G_N$ are
parametrized by some fitting formulae with only one free
parameter $\lambda_1$ that corresponds to the mass scale of the
scalaron.\footnote{The scalaron is a scalar degree of freedom
introduced when the $f(R)$ theory is transformed to the
scalar-tensor theory.} The constraint on $\lambda_1$ obtained
in \cite{Giannantonio:2009gi} is:
\begin{equation}
\lambda_1 < 1900 \, \mbox{Mpc}/h  \; \mbox{ at } \; 95\% \mbox{ c.l.} \, ,
\end{equation}
which then gives the allowed ranges of $\Psi/\Phi$ and
$G_\textrm{eff}/G_N$ for various time and scales $k$. In the
present paper, for the test of $f(R)$ modified gravity we will
simply consider the constraint for the present time ($a=1$) and
the scale $k=0.01 h\mbox{Mpc}^{-1}$:
\begin{eqnarray}
1 \leq & \Psi/\Phi & < 1.996 \, ,
\label{viablepsiphi} \\
1 \leq & G_\textrm{eff}/G_N & < 1.403 \, .
\label{viableGeff}
\end{eqnarray}
This constraint is consistent with GR where both ratios are
unity. We note that according to our experience the test of
$f(R)$ modified gravity is not sensitive to the choice of the
scales. 

We expect the future observations will put more stringent
bounds to $\Psi / \Phi$ and $G_\textrm{eff}/G_N$. For instance,
Guzik \textit{et al.} \cite{Guzik:2009cm} use lensing power
spectra, galaxy-lensing and galaxy-velocity cross spectra to
estimate the ability of the future surveys to test gravity
theories. With $\Lambda$CDM as the fiducial model they give the
following narrower allowed range:
\begin{eqnarray}
0.83 < & \Psi/\Phi & < 1.25 \, ,
\label{futureviablepsiphi} \\
0.84 < & G_\textrm{eff}/G_N & < 1.18 \, .
\label{futureviableGeff}
\end{eqnarray}


\subsection{Solar-system constraints}
\label{intoLG}

The solar-system experiments severely constrain modified
gravity and the deviation from GR, including the $f(R)$ theory,
in the regime with the Ricci scalar $R/H_0^2 \gtrsim 10^5$.
From these local experiments Gu and Lin \cite{Gu:2010} have
obtained a general constraint on $f(R)$ with the chameleon
mechanism:
\begin{eqnarray}
0 \leq & Rf_{RR} & < 0.4 \hspace{1.2em}
\mbox{ as } \quad R/H^2_0 \gtrsim 10^{5} \, ,
\label{GUcondition1} \\
10^{-16} \lesssim & f_{R} & \leq 0 \hspace{2em}
\mbox{ as } \quad R/H^2_0 \sim
10^5 \, . \label{GUcondition2}
\end{eqnarray}


\section{Three Tests of $f(R)$ Modified Gravity}
\label{results}

In this section we test the $f(R)$ theory of modified gravity
via the observations about the physics at three different
scales, ranging from the largest scale to the local scale: (1)
the cosmic background expansion history, (2) the cosmic
structure formation, and (3) the solar-system test of gravity.


\subsection{Cosmic-expansion test}
\label{test-cosmic-expansion}

\begin{figure}
\centering
\includegraphics[width=10cm]{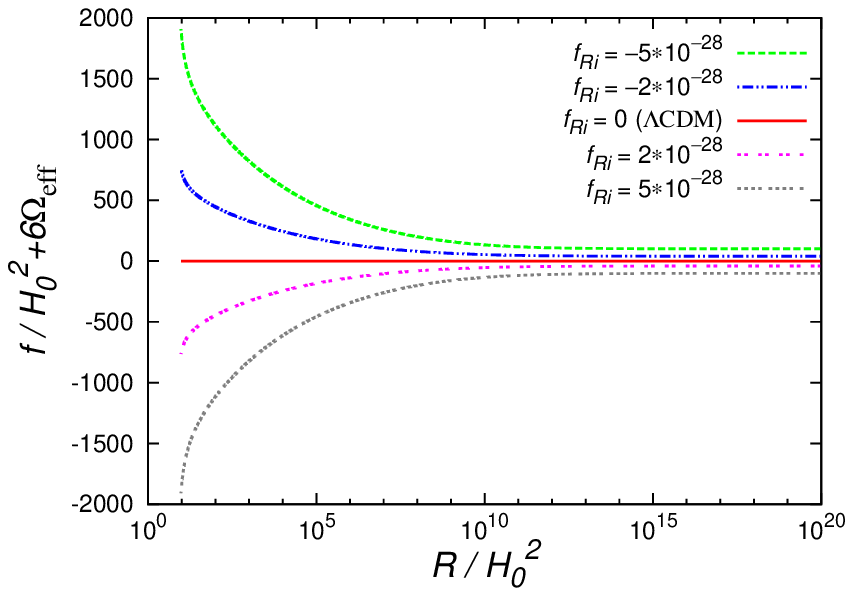}
\caption{Five different designer $f(R)$ w.r.t.\ $w_\textrm{eff}=-1$ for
different initial conditions $f_{Ri}$.}
\label{fvsR}
\end{figure}

To fit the requirement from the observations about the cosmic
expansion history, we consider the designer $f(R)$ constructed
with respect to the given cosmological parameters $\{
\Omega_{m},\Omega_{r},\Omega_\textrm{eff} \}$ consistent with
observations, the effective equation of state $w_\textrm{eff}$
satisfying the constraint in Eqs.\ (\ref{wconst-constraint}) or
(\ref{wCPL-constraint}), and various initial conditions $f_{Ri}
\equiv f_{R}(a_i)$. In particular, we set the parameters
$\Omega_{m} = 0.27$, $\Omega_{r} = 8.4 \times 10^{-5}$,
$\Omega_\textrm{eff} = 1 - \Omega_{m} - \Omega_{r}$ for a flat
universe and the initial time $a_i=10^{-8}$, and consider a
variety of constant $w_\textrm{eff}$ and the CPL-parametrized
effective equations of state,
$w_\textrm{eff}(z)=w_0+w_az/(1+z)$, as follows.
\begin{itemize}
\item For the constant-$w_\textrm{eff}$ case, following the constraint in Eq.\ (\ref{wconst-constraint}) we set $w_a=0$
    and choose $6$ different values for $w_0$:\\
$w_{0}=-1.03$, $-1.00$, $-0.99$, $-0.97$, $-0.95$, $-0.93$. %
\item For the CPL-$w_\textrm{eff}$ case,
following the constraint in Eq.\ (\ref{wCPL-constraint})
we choose four $w_0$ and eight $w_a$:\\
$w_0=-1.04$, $-1.00$, $-0.93$, $-0.85$,\\
$w_a=-1.10$, $-0.72$, $-0.41$, $-0.22$, $0.00$, $0.15$, $0.25$, $0.30$,\\
which together give $32$ different $w_\textrm{eff}$.
\end{itemize}
As an example for demonstration, Fig.\ \ref{fvsR} shows the
designer $f(R)$ with respect to $w_\textrm{eff}=-1$: five
$f(R)$ corresponding to five initial conditions $f_{Ri}$.

We then test these designer models, $f(R;\{w_0,w_a,f_{Ri}\})$,
via the cosmic-structure observations and the solar-system
experiments, as to be shown in the next two subsections.


\subsection{Cosmic-structure test}
\label{test-cosmic-structure}

To test the designer $f(R)$ models via the observations about
the cosmic structure formation, for each model we check whether
the theoretical prediction of the two ratios, $\Psi/\Phi$ and
$G_\textrm{eff}/G_N$, in Eqs.\ (\ref{finalpsiphi}) and
(\ref{finalgeff}) can fit the observational constraint in Eqs.\
(\ref{viablepsiphi}) and (\ref{viableGeff}) for the present
time ($a=1$) and the scale $k=0.01 h\mbox{Mpc}^{-1}$, which is
the only case (time and scale) we consider in the remaining of
the paper. We note that the test of $f(R)$ modified gravity is
not sensitive to the choice of the scales. %

This observational constraint is deduced from
\cite{Giannantonio:2009gi} where the cosmic expansion history
is fitted with the $\Lambda$CDM model. Accordingly, strictly
speaking, it can only constrain the designer $f(R)$ with
respect to $w_\textrm{eff} = -1$. %
Nevertheless, here we also apply this constraint to other
models in order to manifest how various designer $f(R)$ models
might be constrained by the current cosmological observations.



\subsubsection{$1\leq \Psi /\Phi <1.996\,$ for $\, \{a=1,k=0.01h\mbox{Mpc}^{-1}\}$}
\label{resultpsiphi}

\begin{figure}
\centering
\includegraphics[width=10cm]{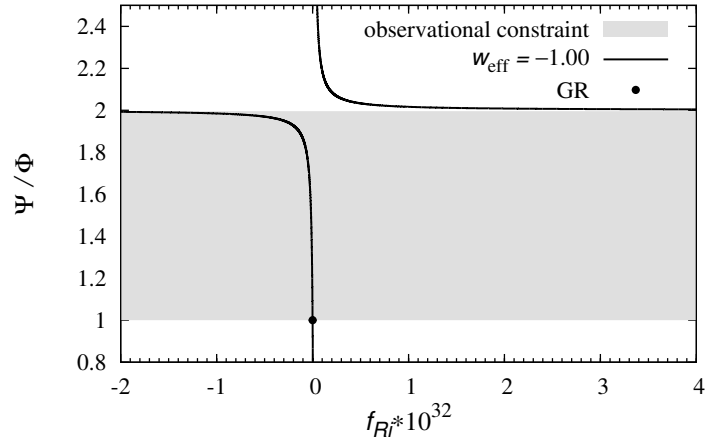}
\caption{The relation of $\Psi/\Phi$ to $f_{Ri}$ for the
$w_\textrm{eff}=-1$ designer $f(R)$ model, as presented by the
solid curve. The grey region shows the observational constraint
of $\Psi/\Phi$, and the dot denotes the GR prediction,
$\Psi/\Phi = 1$.}
\label{LCDM-psiphi}
\end{figure}

\begin{figure}
\centering
\includegraphics[width=10cm]{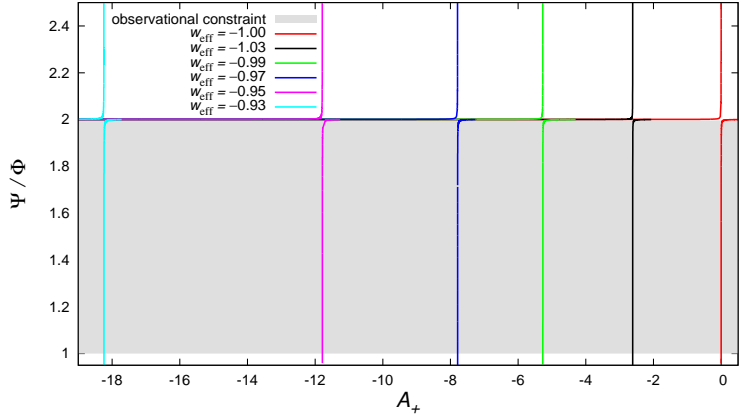}
\caption{The relation of $\Psi/\Phi$ to $A_{+}$ for the designer $f(R)$
models w.r.t.\ $w_\textrm{eff}=-0.93$, $-0.95$, $-0.97$,
$-0.99$, $-1.03$, $-1.00$, as presented by the solid curves from
the left to the right, respectively. The constant $A_{+}$ is related to the the initial condition of $f(R)$ through Eqs.\ (\ref{initialconditionf}) and (\ref{initialconditionfprime}). The grey region shows the observational constraint of $\Psi/\Phi$.}
\label{constw-psiphi}
\end{figure}

\begin{figure}
\centering
\includegraphics[width=10cm]{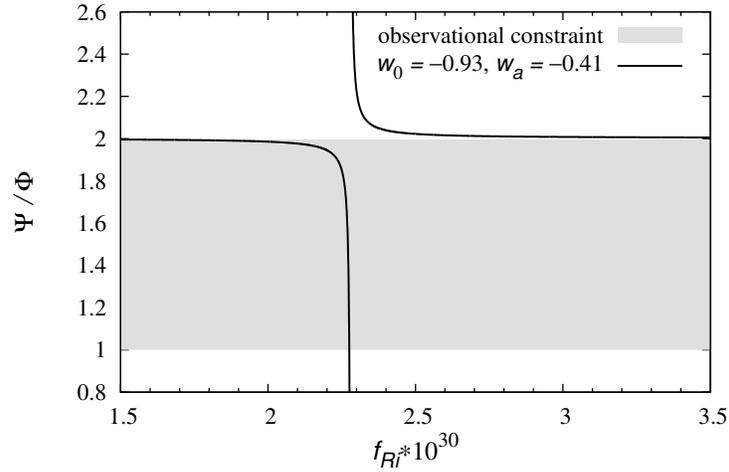}
\caption{The relation of $\Psi/\Phi$ to $f_{Ri}$ for the designer $f(R)$
models w.r.t.\ the best fit of the CPL-parametrized
$w_\textrm{eff}$, i.e., $w_0=-0.93$ and $w_a=-0.41$. The grey
region shows the observational constraint of $\Psi/\Phi$.}
\label{bestfit-psiphi}
\end{figure}

For each $w_\textrm{eff}$ under consideration we explore
numerically the relation of the ratio $\Psi/\Phi
(a=1,k=0.01h\mbox{Mpc}^{-1})$ to the initial condition
$f_{Ri}$. %
As an example for demonstration, in Fig.\ \ref{LCDM-psiphi} we
show the relation of $\Psi/\Phi$ to $f_{Ri}$ for the designer
$f(R)$ model with respect to $w_\textrm{eff}=-1$, as compared
to the observational constraint denoted by the grey region. In
addition, we show in Fig.\ \ref{constw-psiphi} the relation for
various constant $w_\textrm{eff}$, and in Fig.\
\ref{bestfit-psiphi} that for the best fit of the
CPL-parametrized $w_\textrm{eff}$.

We find this relation similar for different $w_\textrm{eff}$:
$\Psi/\Phi \cong 2$ for almost all $f_{Ri}$, while it changes
violently and goes to $\pm \infty$ within a narrow range of
$f_{Ri}$, only in which the designer models may fit the
observational constraint in Eq.\ (\ref{viablepsiphi}). As a
result, one needs to fine-tune the initial condition in order
to have a designer $f(R)$ model consistent with the
observational results about the cosmic structure formation.

\begin{figure}
\centering
\includegraphics[width=10cm]{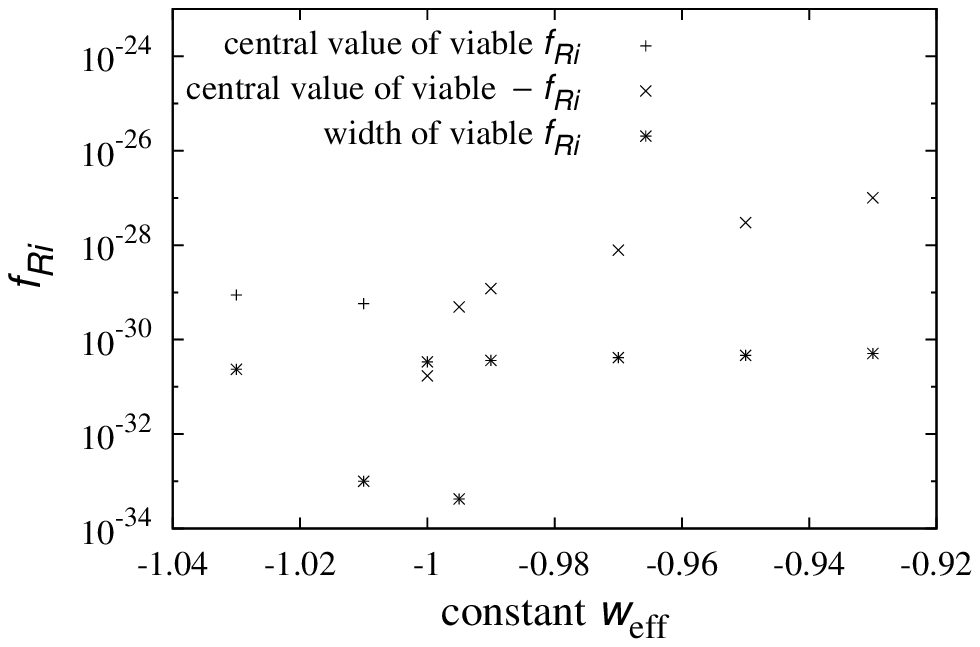}
\caption{The viable initial conditions $f_{Ri}=\bar{f}_{Ri} \pm \Delta
f_{Ri}$ for the constant-$w_\textrm{eff}$ designer $f(R)$
models. The central value $\pm \bar{f}_{Ri}$ is denoted by $+$
or $\times$, and the width $2\Delta f_{Ri}$ of the viable range
denoted by $\ast$.}
\label{constw-viablefRi}
\end{figure}

\begin{figure}
\centering
\includegraphics[width=8cm]{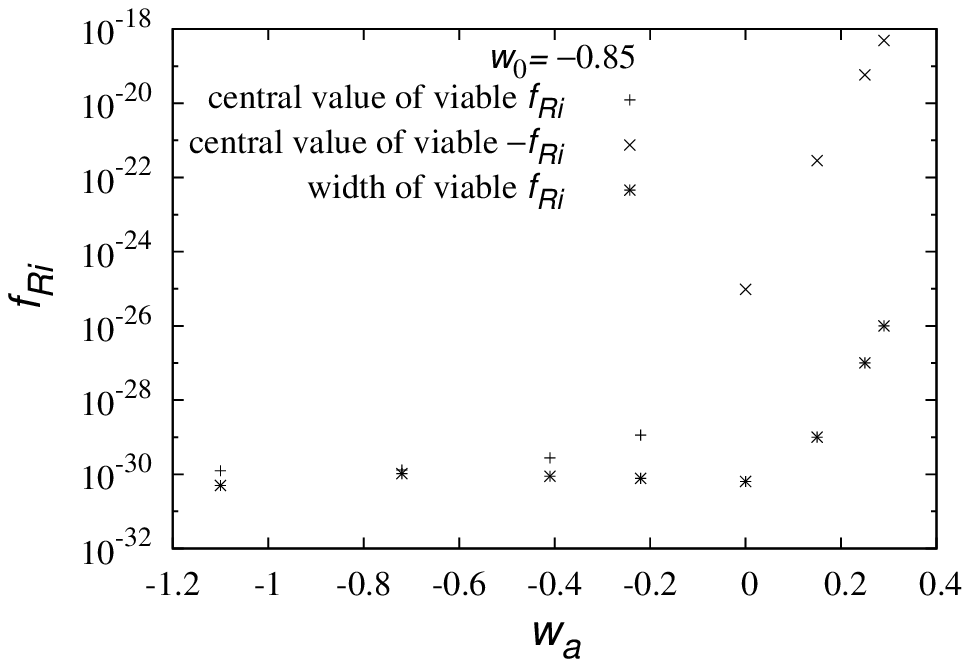}
\includegraphics[width=8cm]{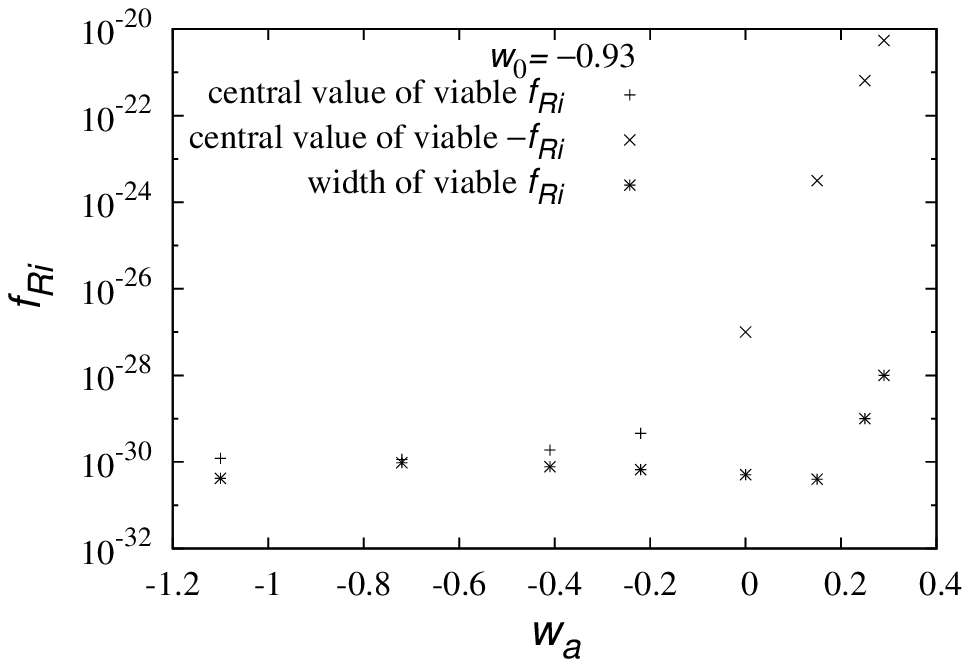}
\includegraphics[width=8cm]{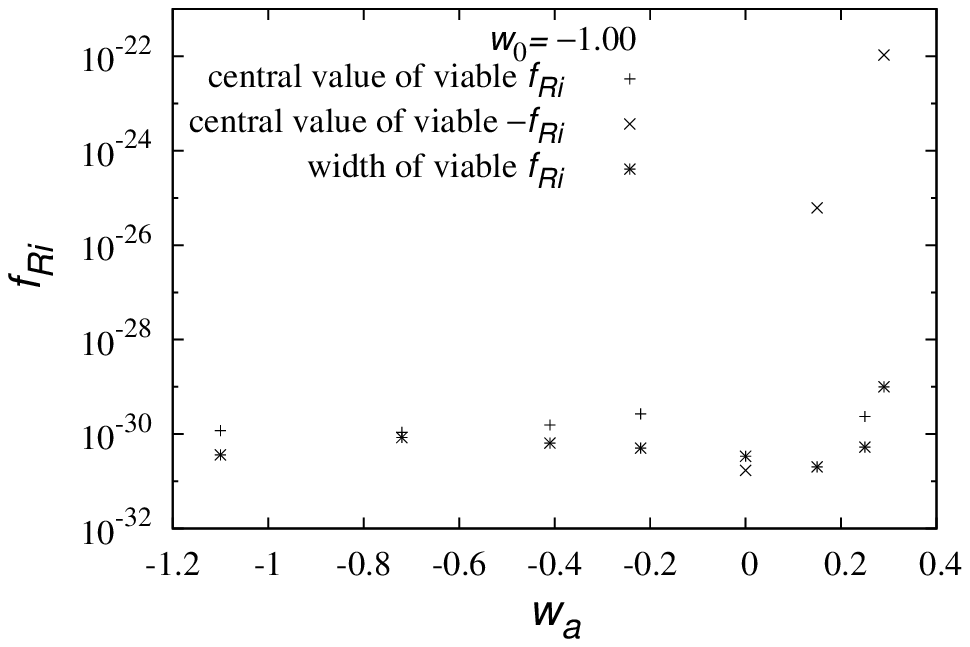}
\includegraphics[width=8cm]{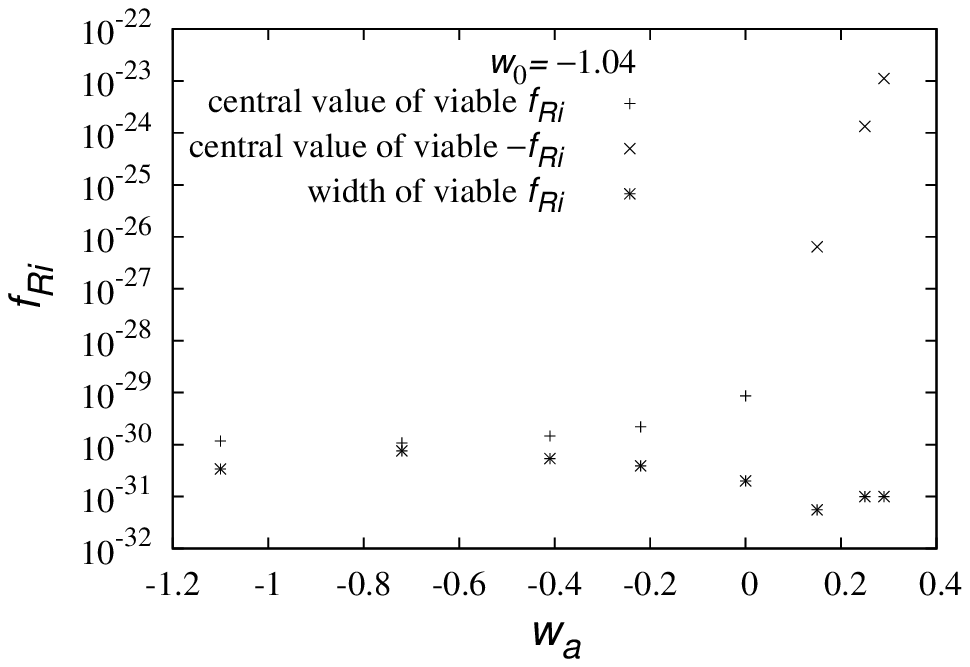}
\caption{The viable initial conditions $f_{Ri}=\bar{f}_{Ri} \pm \Delta
f_{Ri}$ for the CPL-$w_\textrm{eff}$ designer $f(R)$ models.
The central value $\pm \bar{f}_{Ri}$ is denoted by $+$ or
$\times$, and the width $2\Delta f_{Ri}$ of the viable range
denoted by $\ast$.}
\label{CPLw-viablefRi}
\end{figure}

We illustrate the narrow viable range of the initial conditions
$f_{Ri}$ in Figs.\ \ref{constw-viablefRi} and
\ref{CPLw-viablefRi} for the designer $f(R)$ models with
respect to the constant $w_\textrm{eff}$ and the
CPL-parametrized $w_\textrm{eff}$, respectively. We write the
viable range as $f_{Ri} = \bar{f}_{Ri} \pm \Delta f_{Ri}$, and
present in these plots the central value $\pm \bar{f}_{Ri}$
(denoted by $+$ and $\times$) and the width $2\Delta f_{Ri}$
(denoted by $\ast$) of the viable range.


As have been pointed out above, most of the designer $f(R)$
models predict $\Psi/\Phi \cong 2$, which is around the margin
of the current constraint (\ref{viablepsiphi}). We expect the
future cosmological observations will give more detailed and
precise information about cosmic structures, and therefore a
sharper bound on $\Psi/\Phi$. If the future upper bound of
$\Psi/\Phi$ is significantly smaller than $2$, the $f(R)$
modified gravity will be even more fine-tuned and therefore
disfavored. On the other hand, if the future observations give
a bound excluding unity (the GR prediction), GR will be ruled
out.


\subsubsection{$1 \leq G_\textrm{eff}/G_N < 1.403 \,$ for $\, \{a=1,k=0.01h\mbox{Mpc}^{-1}\}$}
\label{resultGeff}

\begin{figure}
\centering
\includegraphics[width=10cm]{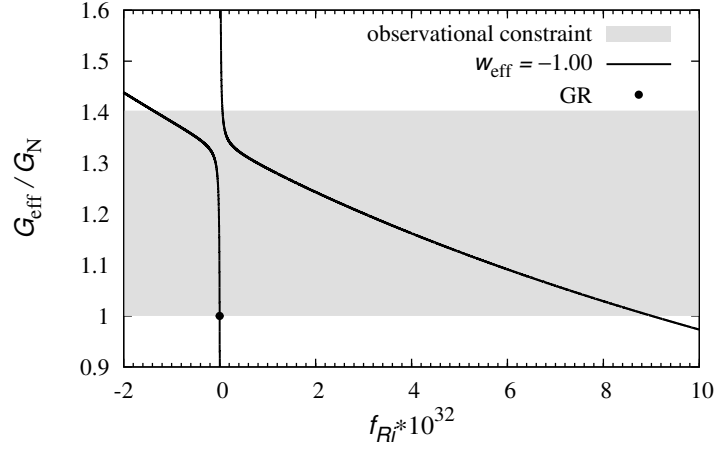}
\caption{The relation of $G_\textrm{eff}/G_N$ to $f_{Ri}$
for the $w_\textrm{eff}=-1$ designer $f(R)$ model, as presented by the solid curve.
The grey region shows the observational constraint of $G_\textrm{eff}/G_N$,
and the dot denotes the GR prediction, $G_\textrm{eff}/G_N = 1$.}
\label{LCDM-Geff}
\end{figure}

\begin{figure}
\centering
\includegraphics[width=10cm]{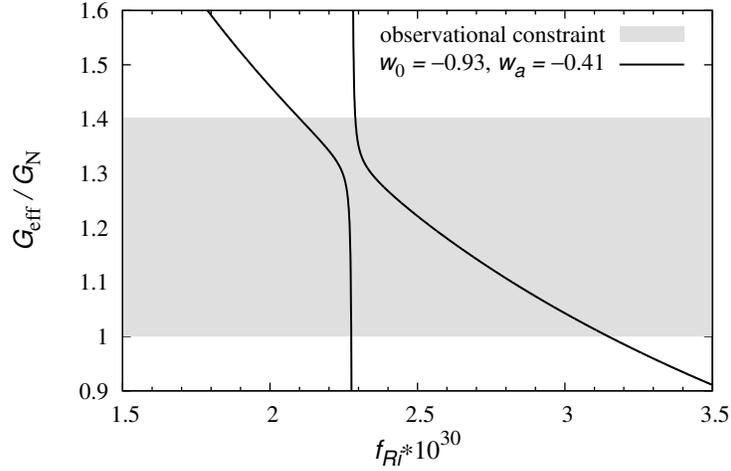}
\caption{The relation of $G_\textrm{eff}/G_N$ to $f_{Ri}$ for the
designer $f(R)$ models w.r.t.\ the best fit of the
CPL-parametrized $w_\textrm{eff}$, i.e., $w_0=-0.93$ and
$w_a=-0.41$. The grey region shows the observational constraint
of $G_\textrm{eff}/G_N$.}
\label{bestfit-Geff}
\end{figure}

For each $w_\textrm{eff}$ under consideration we explore
numerically the relation of the ratio $G_\textrm{eff}/G_N
(a=1,k=0.01h\mbox{Mpc}^{-1})$ to the initial condition
$f_{Ri}$. %
For demonstration, we show in Fig.\ \ref{LCDM-Geff} the
relation for the designer $f(R)$ model with respect to
$w_\textrm{eff}=-1$, and in Fig.\ \ref{bestfit-Geff} that for
the best fit of the CPL-parametrized $w_\textrm{eff}$, as
compared to the observational constraint denoted by the grey
region. We find this relation similar for different
$w_\textrm{eff}$: $G_\textrm{eff}/G_N$ smoothly changes with
$f_{Ri}$ for most of the initial conditions, but varies
violently and goes to $\pm \infty$ within a narrow range of
$f_{Ri}$, around which the designer models can fit the
observational constraint in Eq.\ (\ref{viableGeff}). As shown
in these two figures, there are two separated viable regions of
the initial conditions, one narrow region in the vicinity of
the singularity and the other wider region extended to the
initial conditions away from the singularity. As a result, the
test with the ratio $G_\textrm{eff}/G_N$ gives a weaker
constraint on the designer $f(R)$ models under consideration.



\subsection{Solar-system test}
\label{test-solar-system}

To perform the solar-system test of the designer $f(R)$ models,
for each of the models under consideration we check whether the
requirement in Eqs.\ (\ref{GUcondition1}) and
(\ref{GUcondition2}) is satisfied. We examine the designer
models constructed with respect to the $w_\textrm{eff}$ listed
in Sec.\ \ref{test-cosmic-expansion} and a wide range of the
initial conditions $f_{Ri}$. %
As a result, we find the constraint from the solar-system test
much more stringent:
only the models with $w_\textrm{eff} = -1$ and $f_{Ri} \cong
0$, i.e.\ $f(R) \cong \mbox{constant}$, can fit the constraint. %
These viable models are very close to the case of GR with a
cosmological constant, %
and therefore cannot be differentiated from the $\Lambda$CDM
model, both cosmologically and locally.

\begin{figure}
\centering
\includegraphics[width=8cm]{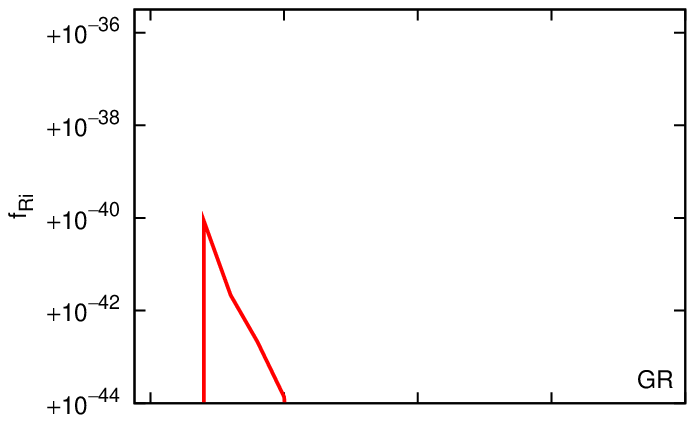}
\hspace{-5.5em}
\includegraphics[width=8cm]{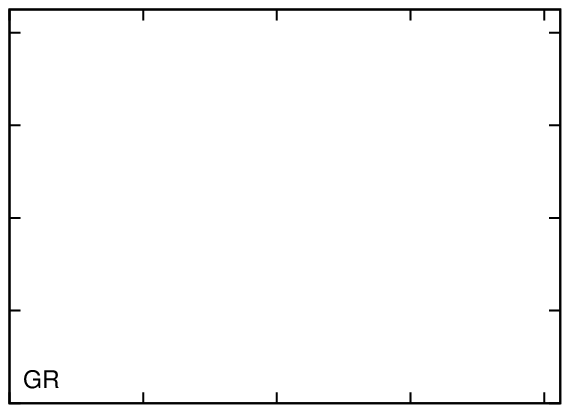} \\
\vspace*{-3em}
\includegraphics[width=8cm]{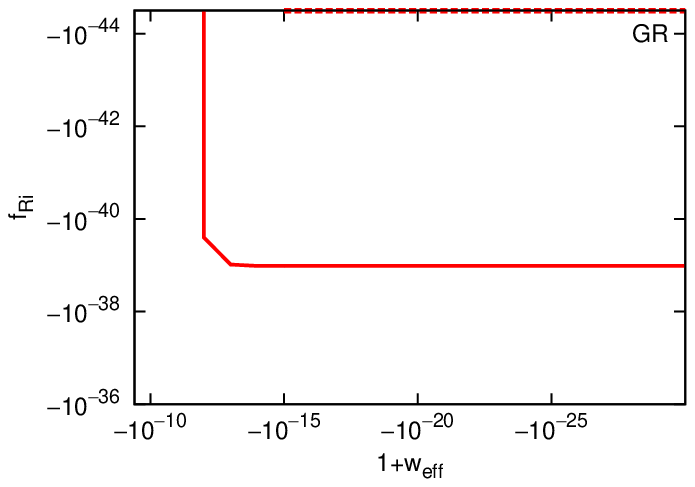}
\hspace{-5.5em}
\includegraphics[width=8cm]{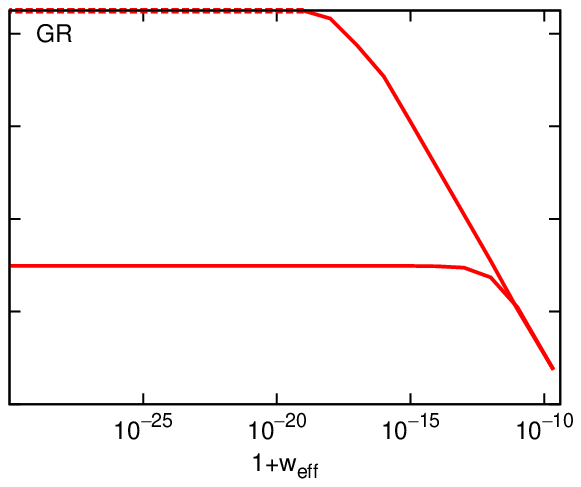}
\caption{The viable $f(R)$ models in the parameter space
$(w_\textrm{eff},f_{Ri})$ for constant $w_\textrm{eff}$ under
the solar-system constraint. The viable region is presented 
by the red contours in the four plots together. The GR point
$(-1,0)$ is between the four plots.}
\label{viableflog}
\end{figure}

To manifest how close to GR the viable models need to be, we
survey in details the small portion of the parameter space $(
w_\textrm{eff},f_{Ri} )$ around the GR point $(-1,0)$ for
constant $w_\textrm{eff}$.
We present the viable region in Fig.\ \ref{viableflog}.
This figure shows that even the cases barely different from GR
are not viable. Thus, among the designer $f(R)$ models under
consideration, only the models closely mimicking GR can pass
the solar-system test. %
Furthermore, we note that the viable models are so close to GR
that in these models not only the cosmic expansion is nearly
$\Lambda$CDM, but also the key quantities for distinguishing GR
from modified gravity in cosmic structures, $\Psi/\Phi$ and
$G_\textrm{eff}/G_N$, agree with the GR prediction (unity)
better than $10^{-6}$.
Accordingly, these models cannot be differentiated from GR via
the cosmological observations, including the observations about
the cosmic structure formation and those about the cosmic
expansion history. As a result, the solar-system test of
gravity rules out the frequently studied $w_\textrm{eff} = -1$
designer $f(R)$ models which are cosmologically
distinct from $\Lambda$CDM.


\section{Conclusion and Discussion}
\label{conclusion}

In this paper we have investigated three tests of the $f(R)$
theory of modified gravity, including two cosmological tests
and one local test. We have invoked the constraints from the
observations of (1) the cosmic expansion history and (2) the
cosmic structure formation and from (3) the solar-system
experiments. We found the constraint from the solar-system test
particularly stringent.

To pass the cosmic-expansion test we have considered the
designer $f(R)$ models constructed with respect to the
constrained effective CPL equation of state $w_\textrm{eff}(z)
= w_0 + w_a z/(1+z)$ and various initial conditions $f_{Ri}
\equiv f_{R}(a=10^{-8})$, and checked whether they pass the
other two tests. We have presented the constraint on $f_{Ri}$
from these two tests for each $w_\textrm{eff}(z)$. We conclude
that the solar-system test rules out all of the designer $f(R)$
models constructed with respect to $w_\textrm{CPL}$, except
those closely mimicking GR with a cosmological constant, i.e.,
the $\Lambda$CDM model. In addition, even passing the
cosmic-structure test alone
would require a fine-tuning of the initial condition of $f(R)$.

We have particularly explored the viable region in the
designer-$f(R)$ parameter space $(w_\textrm{eff},f_{Ri})$
around the GR point $(-1,0)$ for constant $w_\textrm{eff}$,
and found it extremely small. Even a tiny deviation from the GR
point is ruled out by the solar-system test. The designer
models within this viable region are cosmologically
indistinguishable from the $\Lambda$CDM model. More precisely,
their prediction of the cosmic expansion and the cosmic
structure formation agrees with $\Lambda$CDM better than
$10^{-6}$. As a result, the frequently studied
$w_\textrm{eff}=-1$ designer $f(R)$ models that are
cosmologically different from $\Lambda$CDM have been ruled out
by the solar-system test.

We note that despite the stringent constraint from the
solar-system test, it is still possible to construct a viable
$f(R)$ model that is observationally different from GR.
As pointed out by Gu and Lin in \cite{Gu:2010}, the
solar-system test gives a severe constraint [Eqs.\
(\ref{GUcondition1}) and (\ref{GUcondition2})] on the behavior
of the function $f(R)$ for $R/H_0^2$ around or larger than
$10^5$, which corresponds to the time around or before $z \sim
50$. Thus, roughly speaking, the solar-system test and the
observations about CMB and BBN require the deviation from GR be
extremely small, i.e.\ $f(R) \cong \mbox{constant}$, from the
early time to $z \sim 50$, while a significant deviation from
GR is allowed in the present epoch. %
Accordingly, a viable designer $f(R)$ model can be constructed
with respect to the $w_\textrm{eff}$ which is close to $-1$ in
the past but significantly differs from $-1$ recently.
These designer $f(R)$ models are worth consideration in the
quest for a modified gravity model that is cosmologically
distinct from GR with a cosmological constant.


\section*{Acknowledgements}
We thank the Dark Energy Working Group of the Leung Center for
Cosmology and Particle Astrophysics (LeCosPA), particularly
Wolung Lee, Guo-Chin Liu and Huitzu Tu, for the helpful
discussions. %
Lin is supported by the Taiwan National Science Council (NSC)
under Project No.\ NSC 98-2811-M-002-501, %
and Gu by Taiwan NSC under Project No.\ NSC 98-2112-M-002-007-MY3. %
Chen is supported by the Taiwan NSC under Project No.\
NSC97-2112-M-002-026-MY3, by Taiwan's National Center for
Theoretical Sciences (NCTS), and by US Department of Energy
under Contract No.\ DE-AC03-76SF00515.




\clearpage

\clearpage

\end{document}